Title
# A global system of furrows on Ganymede indicative of their creation in a single impact event

Authors
Naoyuki Hirata[1,*], Ryo Suetsugu[2,3], Keiji Ohtsuki[1]

Affiliation
[1] Graduate School of Science, Kobe University, Rokkodai, Kobe 657-8501, Japan
[2] School of Medicine, University of Occupational and Environmental Health, Iseigaoka, Yahata, Kitakyushu, 807-8555, Japan
[3] National Institute of Technology, Oshima College, Suo-Oshima, Yamaguchi, 742-2193, Japan
* Corresponding Author Email address: hirata@tiger.kobe-u.ac.jp

Editorial Correspondence to:
Dr. Naoyuki Hirata
Kobe University
Rokkodai 1-1 657-8501
Tel/Fax +81-7-8803-6566
Email: hirata@tiger.kobe-u.ac.jp

Highlight
- ■ We studied the global distribution of ancient tectonic troughs, furrows on Ganymede
- ■ We found that furrows used to be a global-scale multi-ring structure.
- ■ If it is impact origin, an 150-km radius impactor is plausible

Abstract

Furrows are a concentric system of tectonic troughs, and are the oldest recognizable surface feature on Ganymede. We analyzed the distribution of furrows utilizing Voyager and Galileo images and found that furrows over Ganymede's surface are part of a global concentric circular structure. If this multi-ring structure is impact origin, this is the largest impact structure identified so far in the solar system. Deviations of the shapes of the furrows from the concentricity are small everywhere, which implies that the relative location of the blocks of the dark terrains over the entire surface of Ganymede has not changed appreciably even during formation of the bright terrains. The estimate of the impactor size is difficult, but an 150km-radius impactor is consistent with the observed properties of furrows. The furrow-forming impact should have significant effects on the satellite's geological and internal evolution, which are expected to be confirmed by future explorations of Jupiter's icy moons, such as the JUICE (Jupiter Icy moon Explorer) or Europa Clipper mission.

# 1. Introduction

Several spacecraft such as Voyager and Galileo spacecraft and ground-based observations revealed Ganymede's remarkable properties. One of the most notable surface structures on Ganymede is a concentric system of tectonic troughs, termed furrows (Fig. 1AB). Furrows exist only on geologically old terrains, termed dark terrains. The dark terrains, consisting of 33-34 % of Ganymede's total surface, are heavily cratered and well preserve many ancient craters, as on Callisto (Frieden and Swindell, 1976; Pappalardo et al., 2004; Passey and Shoemaker, 1982; Shoemaker and Wolfe, 1982; Smith et al., 1979a; Smith et al., 1979b). Because furrows are crosscut by any recognizable impact craters exceeding 10 km in diameter, they are regarded as the oldest recognizable surface features on Ganymede (Passey and Shoemaker, 1982). The rest of Ganymede's surface, termed bright terrains, is young terrains that have numerous tectonic features called grooves and lack furrows.

It was proposed that some of the furrows are fragments of multi-ring impact basin structures similar to the Valhalla or Asgard basins on Callisto (Fig 1C), which are thought to be formed by collapse of a crater basin when the excavation depth is comparable to the thickness of the lithosphere of the satellite (McKinnon and Melosh, 1980; Melosh, 1982; Smith et al., 1979a). According to Voyager-era studies, the largest one exists across Galileo Regio and Marius Regio (so-called the Galileo-Marius furrow system) and it has at least hemispherical scale (Schenk and McKinnon, 1987; Zuber and Parmentier, 1984). On the other hand, Voyager images did not cover all of the dark terrain blocks, such as Galileo, Perrine, or Nicholson Regio. Therefore, they do not constrain the scale of Galileo-Marius system. Galileo images covered the regions not imaged by Voyagers and showed that the dark terrain is ubiquitously distributed over the surface of Ganymede and most of them have furrows. However, there are still some areas of dark terrain that are not sufficiently imaged by either mission. For example, few furrows have been found in Melotte Regio, which is presumably because the resolution, illumination, and/or emission of images were not suitable for detecting furrows yet. Future remote-sensing observations may find many more furrows in the dark terrains.

In the present work, first we reanalyze the distribution of furrows utilizing both Voyager and Galileo images, and update our knowledge about the scale of this structure (Section 2). We show that this is a structure not in a hemispherical scale but in a global scale, and that it would have been formed by a single large impact. In addition, we estimate the impactor size based on the model of the scale of furrows and numerical simulation (Section 3). We discuss implications of our findings for the evolution of the satellite in Section 4, and our conclusions are summarized in Section 5.

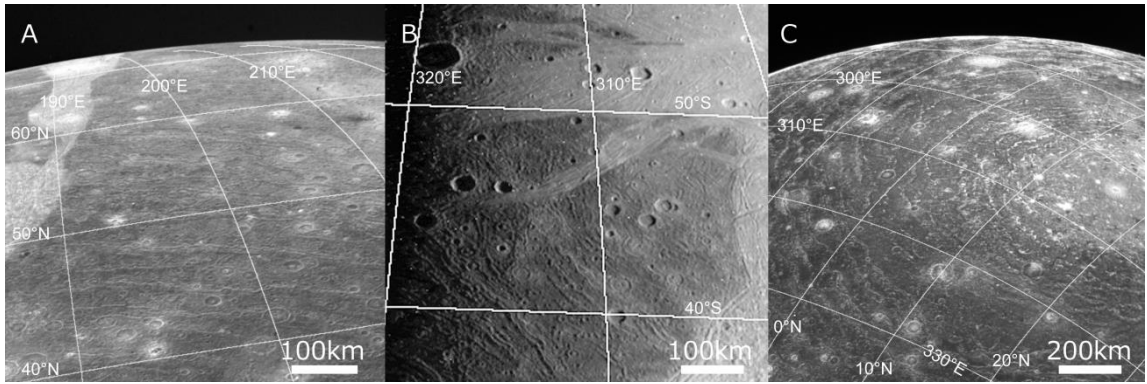

**Fig. 1.** Furrows on Ganymede and the Valhalla ring system on Callisto. (A) A concentric system of tectonic troughs in Galileo Regio of Ganymede, termed furrows (C2063641). (B) A parallel system of furrows in Nicholson Regio of Ganymede (s0401701078), which are one of the furrows newly identified from the Galileo images. Although this region was imaged by Voyagers, the lighting and emission angles of the Voyager images were not suitable for detecting furrows. (C) Valhalla basin on Callisto and its multi-ring system (C1642151).

## 2. The scale of the furrow system
### 2.1. Previous studies

The Voyager 1 spacecraft has imaged the sub-jovian hemisphere of Ganymede at a resolution of up to 1.0 km/pixel and Voyager 2 has imaged the anti-jovian hemisphere at a resolution of up to 500 m/pixel. Utilizing these images, furrows are identified in Galileo, Marius, Barnard Regio, the northeastern portion of Perrine Regio, and the eastern portion of Nicholson Regio (e.g. Schenk and McKinnon, 1987; Zuber and Parmentier, 1984). Examination of Voyager images shows that the largest furrow system have a concentric patterns centered at 20°S 180°W across Galileo Regio and Marius Regio, whose scale is at least hemispherical (Fig. 2A) (Schenk and McKinnon, 1987; Zuber and Parmentier, 1984). Because of poor visibility of Voyager images for some area, the scale of this furrow system was uncertain.

The Galileo spacecraft has imaged many portions at moderate resolution (~ 0.8-3.6 km/pixel) during the six close encounters (Pappalardo et al., 2004). Galileo has imaged many of the dark terrain not resolved by Voyagers. The global geologic map of Ganymede, representing a synthesis of our understanding of Ganymede geology after the conclusion of the Galileo mission, has been published (Collins et al., 2013). Collins et al. (2013) mapped newly-identified furrows in far eastern Galileo Regio, western portion of Perrine Regio, Melotte Regio, the western portion of Nicholson Regio, and unnamed blocks near Ta-urt craters, although the distribution of the furrows is not discussed then.

## 2.2. Our observation

We utilized Voyager/Galileo images, a global mosaic (U.S. Geological Survey, 2003), and two previous measurements of furrows (Collins et al., 2013; Schenk and McKinnon, 1987), and analyzed the distribution of furrows on Ganymede. We used ISIS3 software produced by the U.S. Geological Survey for both map projections and radio-metrically calibrations. We did not add new furrows in this work. We developed azimuthal equidistant projection maps centered at 20°S 180°W (Fig. 2A), and its exactly opposite hemisphere (Fig 2B). As shown in Fig. 2A, we confirmed that the Galileo-Marius furrow system is aligned in concentric circles centered at a single point, 20°S 180°W. Interestingly, we found that not only furrows in Galileo and Marius Regio but also ones even on the opposite hemisphere are aligned in concentric circles centered at the single point. Although previous works estimated that the system has at least a hemispherical scale (Schenk and McKinnon, 1987), we noticed that furrows over Ganymede form a global concentric system and Galileo-Marius system is a part of the global system. In other words, the scale of the Galileo-Marius furrow system is much greater than previously supposed. Although the dark terrain remained in only one-third of the current surface, our finding indicates that Ganymede used to have a global-scale multi-ring system formed by a single event before the formation of the bright terrain. If these furrows are impact origin, this is the largest impact structure identified so far in the solar system. In order to see the deviation of the shapes of the furrows from the concentricity, we developed an oblique simple cylindrical projection (Fig. 3 and 4). Fig. 3 shows the dark terrain projected on the oblique simple cylinder, and the vertical direction indicates the equidistance from the center. Fig. 4 shows clops of transects of the dark terrains shown in Fig. 3A. Figs. 3 and 4 indicate that the furrows seem to be aligned in the horizontal direction everywhere.

In order to assess the above arguments quantitatively, we geometrically analyzed the furrow system over Ganymede following the method of Schenk and McKinnon (1987). The details are described in Appendix A. As a result, we estimated the geometric center of the furrow system to be 21.3°S 179.4°W using furrows in Galileo and Marius Regio and 22.9°S 178.4°W using furrows over Ganymede, respectively. These geometric centers obtained in the present work are consistent with the one obtained by Schenk and McKinnon (1987). Furthermore, we calculated the deviation from concentricity of each furrow segment and made its histograms, assuming the center to be 21.3°S 179.4°W (Fig. 5). The histograms for the furrow system over Ganymede showed a sharp peak at 0-5° deviation from concentricity, with 85% of the ring segments aligned within 30° of concentricity (Fig. 5a). According to Schenk and McKinnon (1987), the histogram for Valhalla ring system showed a sharp peak at 0-10° deviation from concentricity, with 90% of the ring segments aligned within 30° of concentricity. Thus, the concentricity of the furrow system over Ganymede was found to be comparable to that of Valhalla ring system. Note that the histograms for Ganymede's furrow system did not exclude radial and sub-radial furrows, but ones for Valhalla ring system done by Schenk and McKinnon (1987) excluded them. Similar plots for furrows in the opposite hemisphere (Nicholson, Perrine, and Melotte Regios) also showed a sharp peak at 0-10° deviation from concentricity, with 90% of the ring segments aligned within 30° of concentricity (Figs. 5b and 5c). Therefore, the concentricity of the furrow system over Ganymede is sufficiently small to be regarded as a single concentric circle system. It has been proposed that furrows lying partly in Galileo Regio shown in the plate G5 in Fig. 3 (and Fig. 4F) form a distinct small multi-ring system (Prockter et al., 2002). However, this G5 system is also sufficiently aligned with the global-scale furrow system (Fig. 5d), and therefore, it is possible that the G5 system could be regarded as a part of the global-scale furrow system (this does not mean that the G5 system is not distinct multi-ring system).

It is known that Marius Regio (especially, the region within the colatitude of 30° from 20°S 180°W) has few furrows (Fig. 6) (Schenk and McKinnon, 1987; Zuber and Parmentier, 1984). The region within 30 degrees from the center of the above furrow system ($r < 1380$ km, where $r$ is the radial distance from the center on the satellite's surface) may be the structure corresponding to the central zone without trough or ridge of the Valhalla basin ($r < 360$ km). This furrow system extends in Barnard Regio (the region from the colatitude 163 to 170 degree), which implies that the furrow system on Ganymede has the extent of 1380 km $< r <$ 7800 km. Unlike the Valhalla ring system, the furrow system of Ganymede does not have either obvious ridged zone or outward-facing normal faults in any dark terrains, which has been already pointed out by McKinnon and Parmentier (1986) based on Voyager images, and our update based on Galileo images has confirmed it again.

There is a possibility that a true impact site is not the Marius side (20°S 180°W) but Barnard side (20°N 0°W). At the moment, we cannot argue conclusively which side is the true center. That said, furrows on the Marius side are typically much more obvious than those in the Barnard side. Also, if the true center is on the Barnard side, the central region without furrows is too small. Therefore, in this work, we assume that the true center of the system is on the Marius side.

This implies that the relative locations of the blocks of the dark terrains did not change appreciably. Previous work shows that deviations of the shapes of the furrows in Marius Regio and Galileo Regio from the concentricity are small, and therefore, lateral movement of each dark terrain block does not occur between Marius Regio and Galileo Regio (Schenk and McKinnon, 1987). Our observation indicates that relative locations of each dark terrain block over the entire surface of Ganymede (not only Marius and Galileo but also Perrine Melotte, Nicholson, and Barnard Regio) have been unmodified since the formation of the furrows, despite of their great distances. We verified quantitatively how large lateral movement between each dark terrain block can be detected by the concentricity of furrows. As an example, we shifted the western portion of Nicholson Regio (N2 in Fig. 3) slightly, and examined how the histogram in Fig. 5c changes. Results are shown in Fig. 7 (see Appendix A). In the case of 10° movement, the majority of the deviation was within 20 degrees, but the peak positions were slightly shifted. In the case of 20° movement, the peak positions were shifted or separated. In the case of 30° movement, the histogram no longer followed a normal distribution. Note that the original histogram (Fig. 5c) showed a more ideal normal distribution curve peaked at zero degree than these 15 examples shown in Fig. 7. Therefore, it is likely that the lateral movement of the relative position between Nicholson and Galileo-Marius system would be 20 degrees at maximum (this corresponds to 920 km on Ganymede).

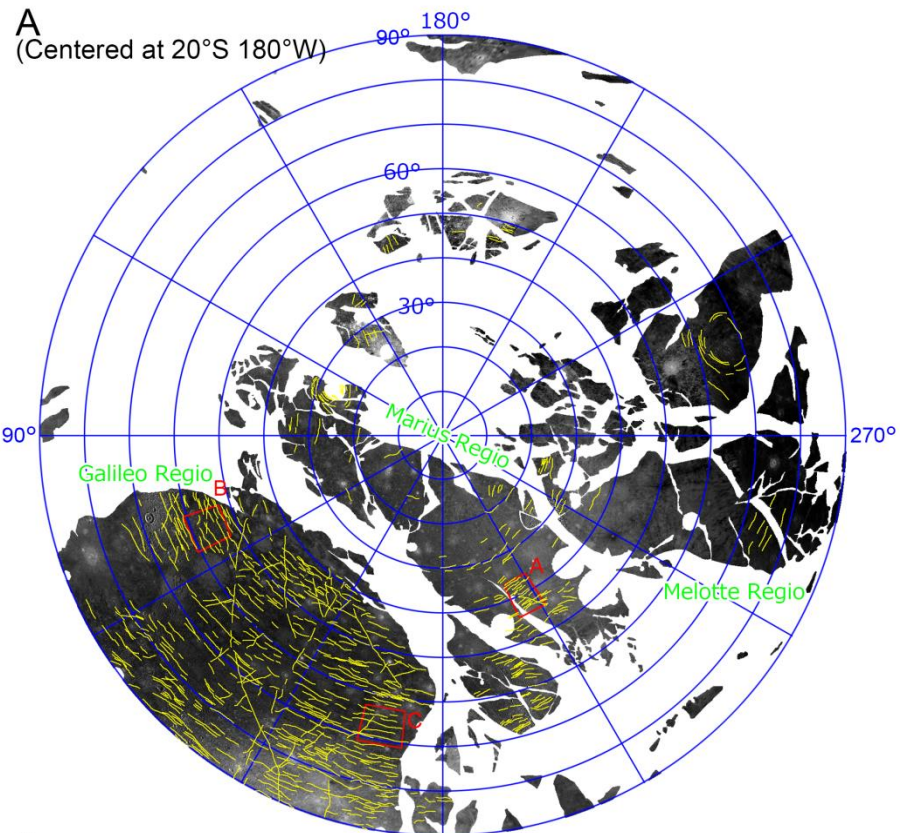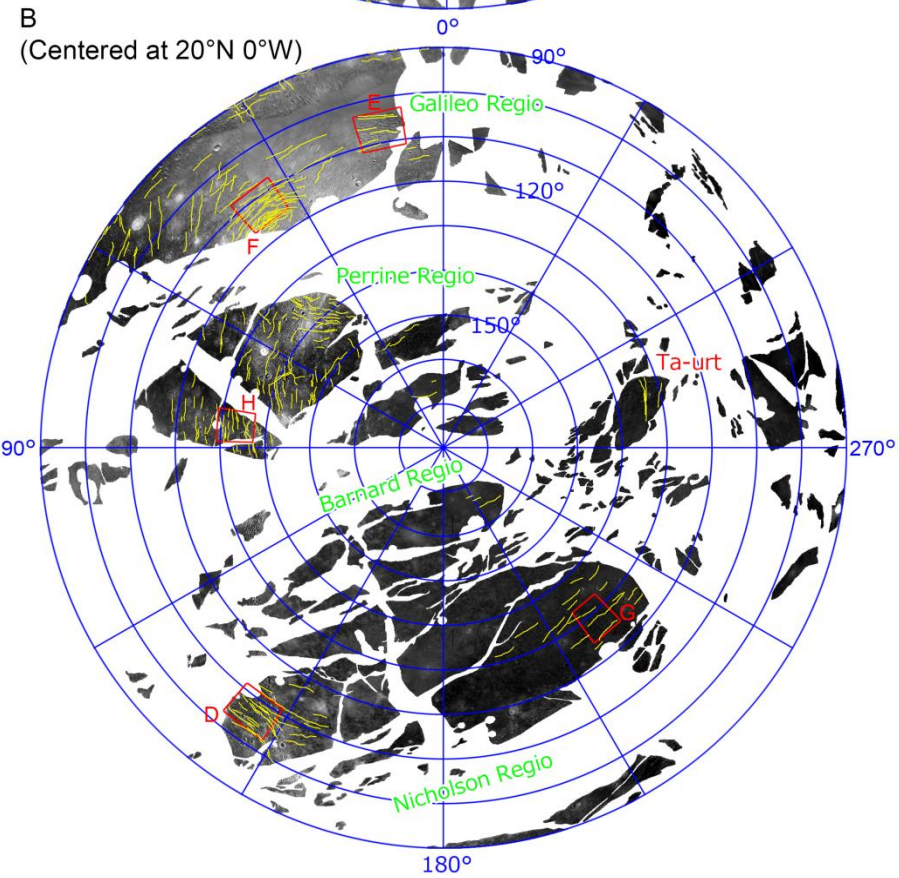

Fig. 2. Distribution of furrows (yellow lines) projected in azimuthal equidistant map centered at 20°S 180°W (plate A) and its exactly opposite point, 20°N, 0°W (plate B). White regions indicate the bright terrains. The distributions of the dark terrain and furrows are from Collins et al. (2013) and Schenk and McKinnon (1987). Red rectangles show the location of each plate in Fig. 4.

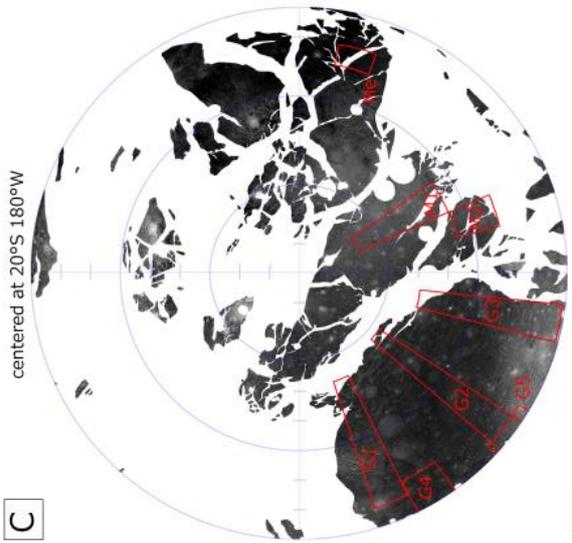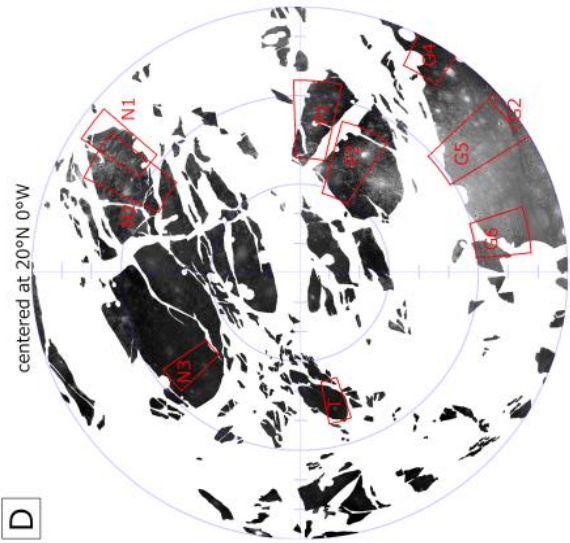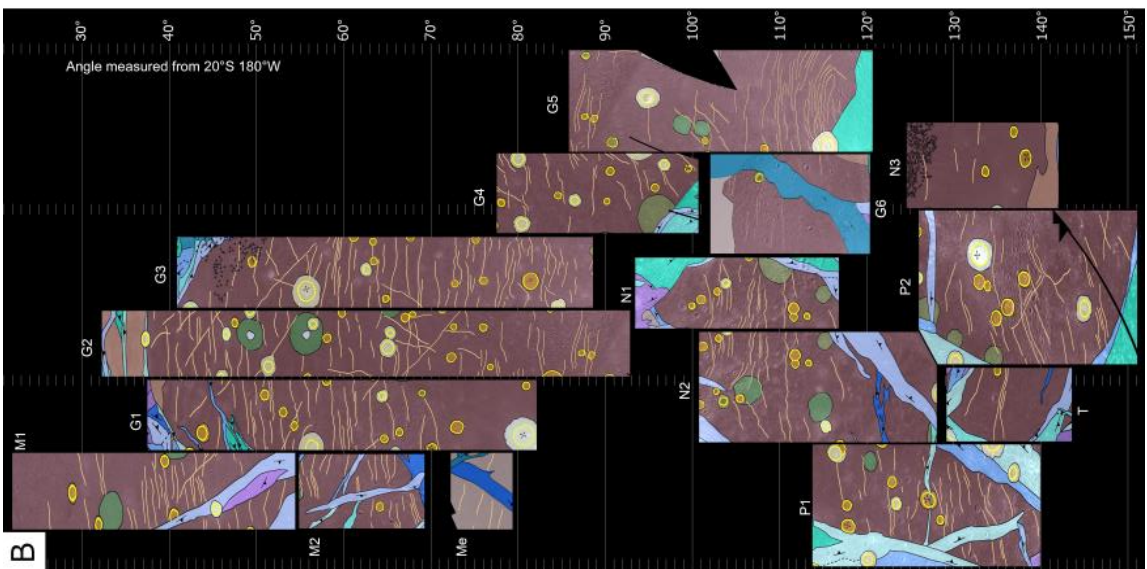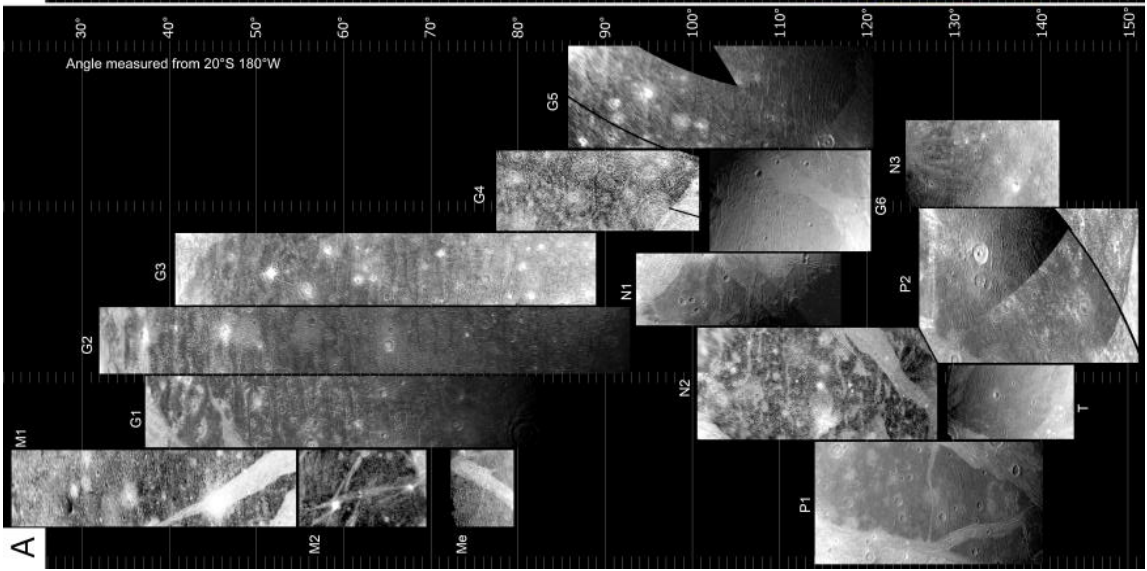

**Fig. 3.** Transects of the dark terrain of Ganymede along a radial direction from 20°S 180°W to its antipodal point. Vertical axes in plate (A) and (B) represent the colatitude measured from 20°S 180°W. Note that 10° is equivalent to 460 km. Maps in (A) and (B) are shown in oblique simple cylindrical projection poled at 20°S 180°W. Plate (B) is modified from the global geologic map (Collins et al., 2013). Yellow lines and brown areas mean the furrows and the dark terrains identified by Collins et al. (2013). The location of each transect is shown in (C) and (D), which are the projected on the azimuthal equidistant maps centered at 20°S 180°W and its exactly opposite point, 20°N 0°W, respectively. The original-sized image of this figure is uploaded as supplementary Figure S1.

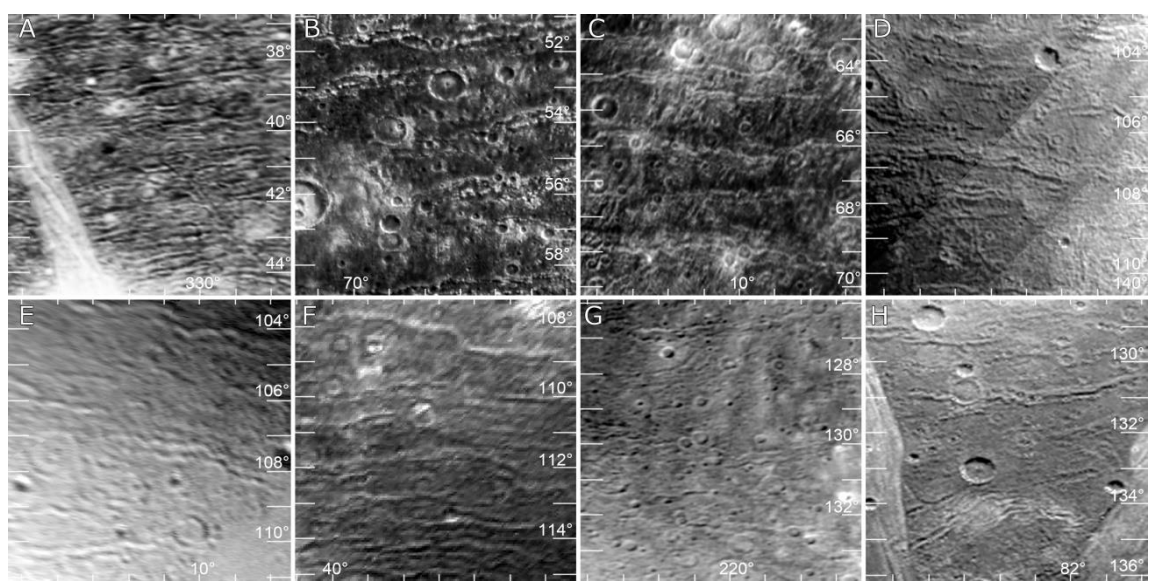

**Fig. 4.** Close up views of the furrows. Vertical axis represents the colatitude measured from 20°S 180°W. The oblique projection is utilized for this image, and the horizontal axis represents longitude of the oblique simple cylinder. In this figure, the fact that furrows are aligned with the horizontal direction means that they are aligned with concentric circles centered at 20°S 180°W. These images shows Marius Regio (A), Galileo Regio (B, C, E, F), Nicholson Regio (D, G), and Perrine Regio (H), respectively. The detailed locations of each plate are shown in red rectangles of Fig. 2. The appearance of the furrows in each plate (especially, BCDEFH) is similar to each other despite of great difference in their distance, which indicates all have a same origin.

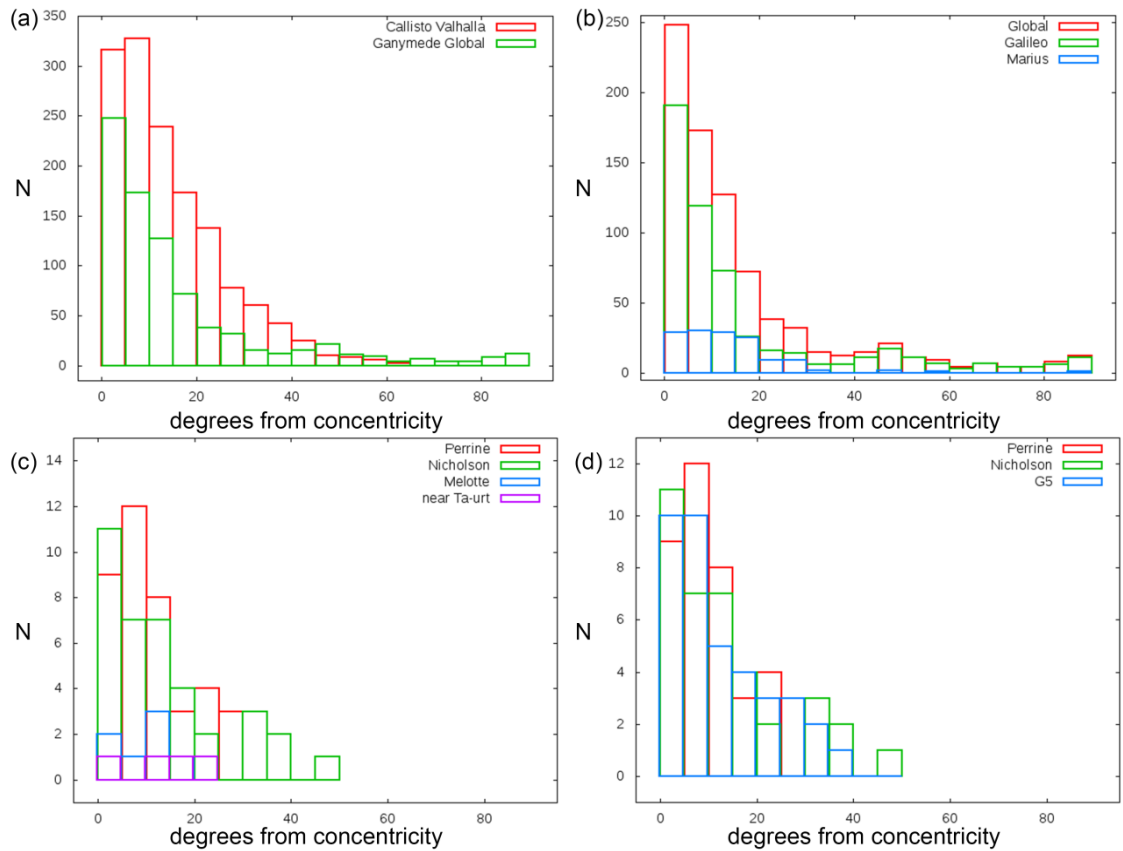

Figure 5. Histograms of the deviation from concentricity (angle of intersection between furrow segments and concentric circles centered at 21.3°S 179.4°W). Vertical axis means the number of furrow segments and horizontal axis the degree from concentricity (0° means segments are aligned with concentric circles, while 90° means those are radial to them). Histogram for (a) Valhalla ring system from Schenk and McKinnon (1987) and Ganymede's furrow system over Ganymede from our measurement, (b) Ganymede's furrow system over Ganymede, in Galileo Regio, and in Marius Regio, (c) for Perrine, Nicholson, Melotte, and unnamed blocks of the dark terrains near Ta-urt crater, and (d) for the G5 small multi-ring system. See also Appendix A.

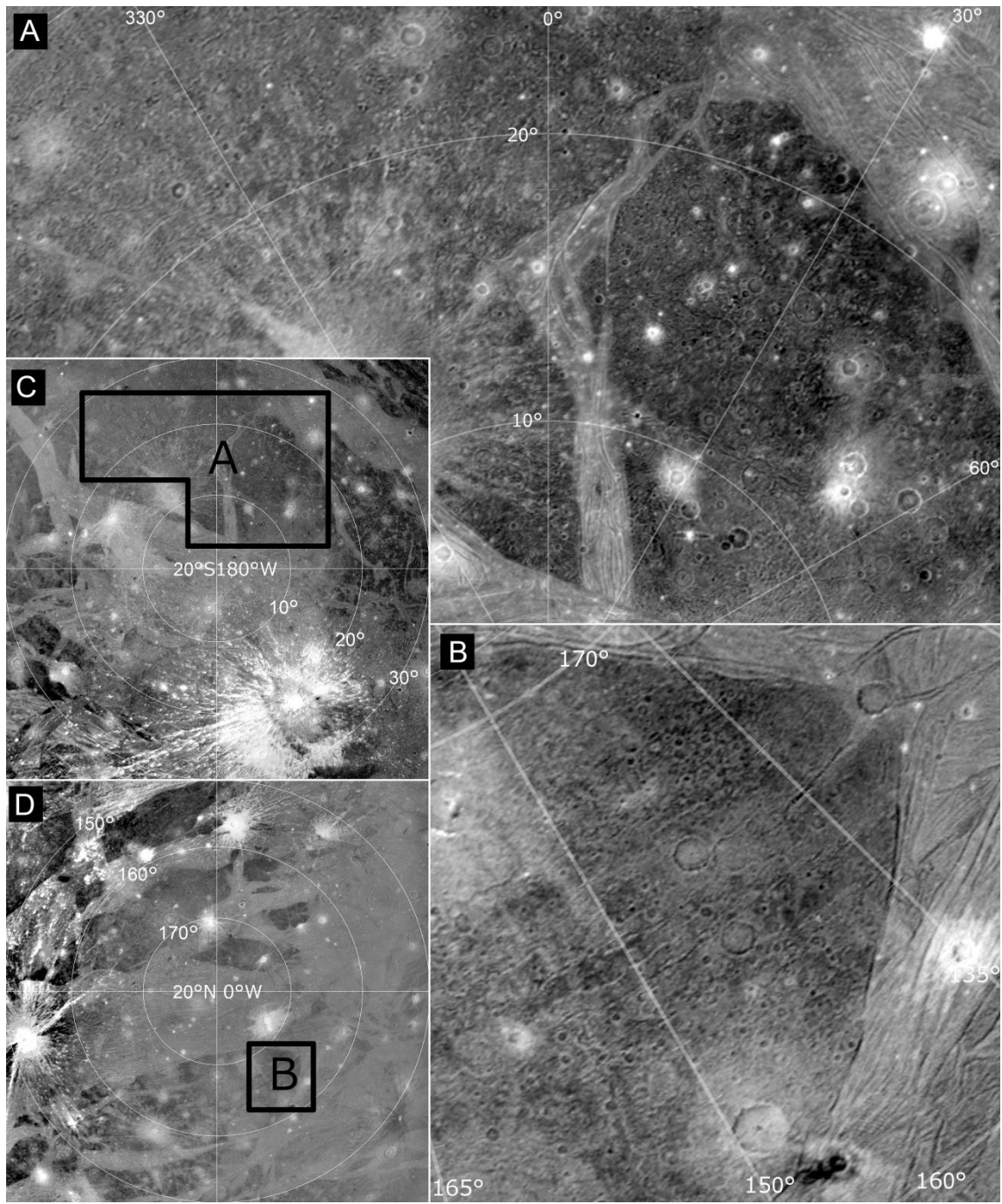

**Fig. 6.** (A) Close up view of the dark terrains near 20°S 180°W (Marius Regio). (B) That near 20°N 0°W (Barnard Regio). (C) and (D) show the locations of (A) and (B).

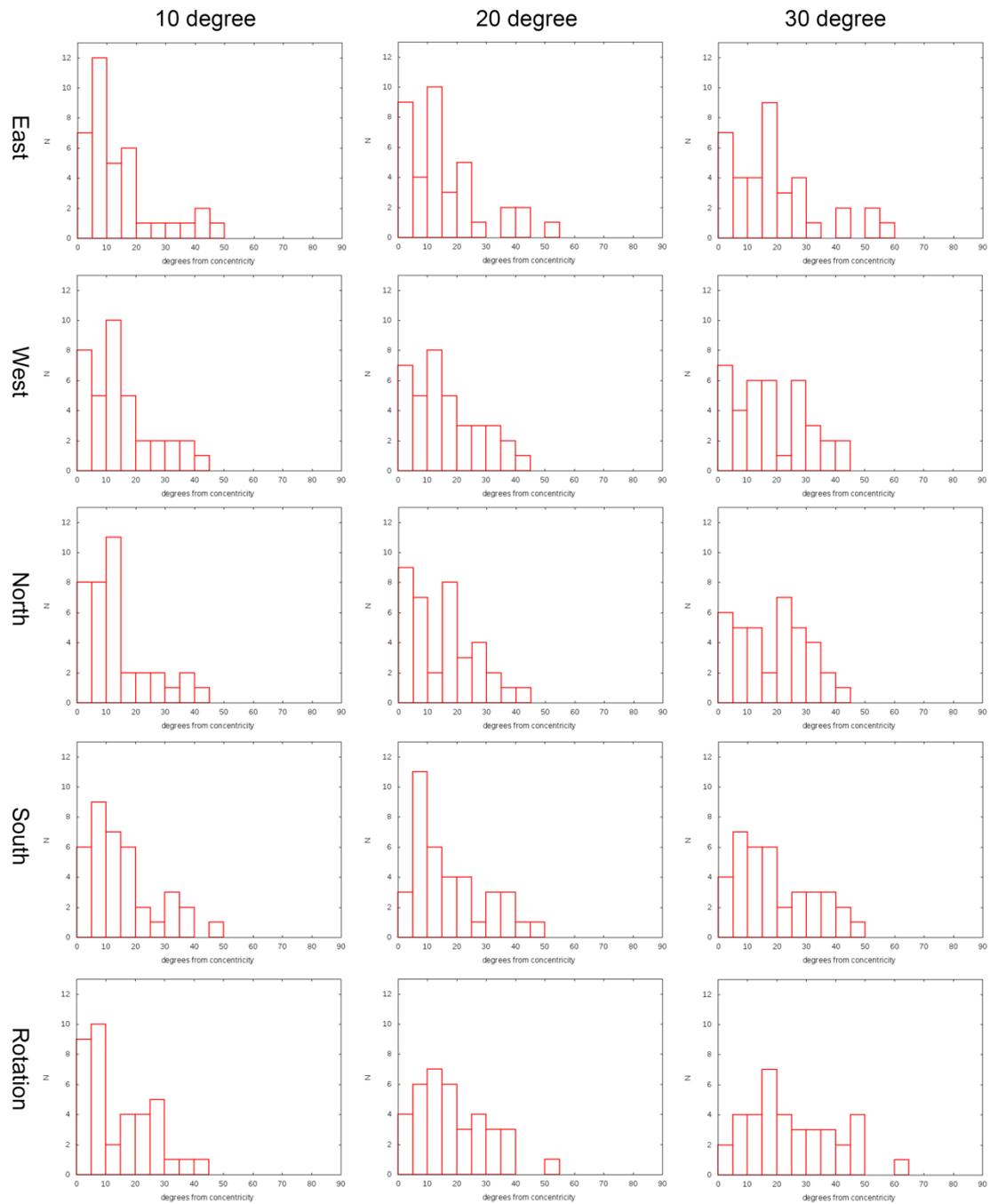

**Figure 7.** The histogram of the deviation from concentricity between furrow segments in Nicholson Regio that were moved slightly away from its original location. We show 15 examples: 10°, 20°, and 30° movement toward north, south, west, and east and 10°, 20°, and 30° rotation at the current location without shift. The original one is shown in Fig. 5c. See also Appendix A.

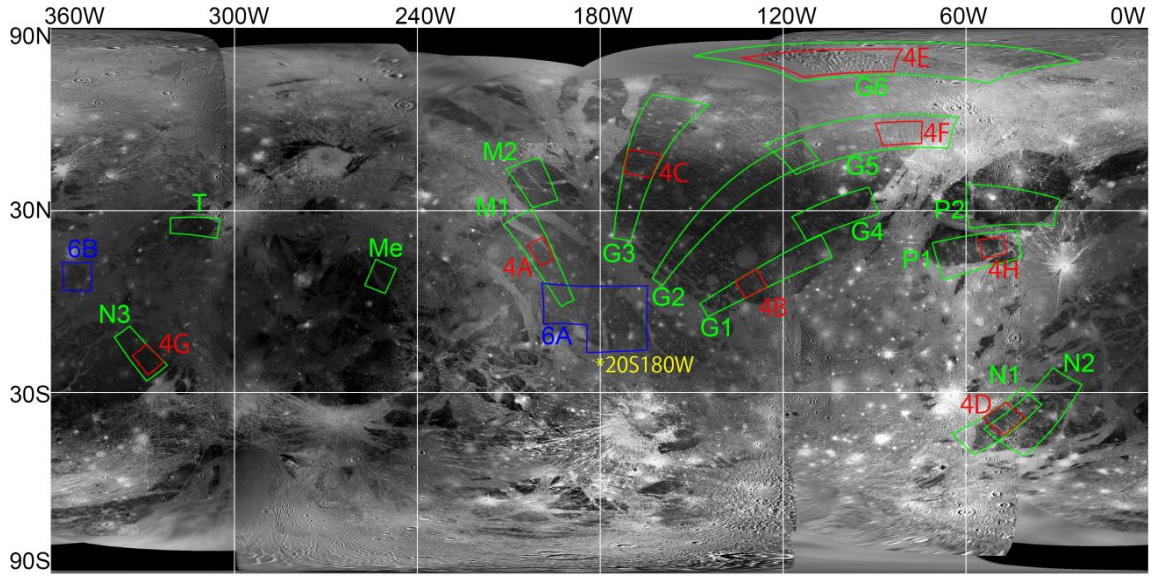

**Figure 8**. The location of each plate in Figs. 3 (green), 4 (red), and 6 (blue) in simple cylindrical projection map of Ganymede.

## 3. Impactor size

Estimating the size of the furrow-forming impactor (FFI) is complicated because of the absence of an identifiable clear rim. Absence of clear rim around a seemingly impact-generated structure is not unusual; for example, the Valhalla basin or Asgard basin on Callisto, which are thought to have been formed by an impact, also lacks such a rim (Schenk, 1995). We attempt to estimate the size of FFI based on three considerations: (i) Comparison with Valhalla and Asgard basins on Callisto and palimpsests on Ganymede, (ii) a thin plastic lithosphere model and the extent of furrows, and (iii) numerical simulation.

### 3.1. Comparison with Valhalla and Asgard basins on Callisto and palimpsests on Ganymede

Valhalla and Asgard are representative multi-ring impact basins on Callisto. Valhalla basin can be divided into three distinct zones outward from the center: a central smooth zone ($r < 360$ km, where $r$ means the radial great-circular distance from the center of the basin), an inner ridge and trough zone ($360 \text{km} < r < 950$ km), and an outer graben zone ($950 \text{ km} < r < 1900$ km). Asgard has a central smooth zone ($r < 100$ km), an inner ridge zone ($100 \text{ km} < r < 400 \text{km}$), and an outer graben zone ($400 \text{km} < r < 940 \text{km}$)

(Greeley et al., 2000; Schenk, 1995). It is known that the crater density in the central palimpsest and inner ridge zone of the Valhalla basin is 3.5 times lower than on the adjacent unmodified cratered terrain, while the crater density in the outer graben zone is intermediate between the inner zones and unmodified cratered terrain and apparently increasing linearly outward (Passey and Shoemaker, 1982; Schenk, 1995). The former is due to complete obliteration and tectonic disruption of the original surface, and the latter due to obliteration by continuous ejecta blanket (Schenk, 1995). Although these crater rims collapsed or did not form, these original crater diameters were estimated to be equivalent to 1000 km for Valhalla and 675 km for Asgard based on the mapping of ejecta and secondaries (Schenk and Ridolfi, 2002). If so, we can estimate Valhalla-forming impactor radius to be $R_{imp} \approx 50$ km, using the simple scaling law shown in Eqs. (5) and (6) in Zahnle et al. (2003), 15 km/s as the impact velocity onto Callisto (Zahnle et al., 2003), 2000 kg/m$^3$ as the density of the impactor (equivalent to Ceres or Pluto), and 900 kg/m$^3$ as the density of Callisto's crust.

     A palimpsest is a circular albedo feature around an ancient crater seen on Ganymede and Callisto (Smith et al., 1979a). Palimpsests on Ganymede, such as Memphis Facula, can be divided into four distinct zones outward from the center: the central plains (*cp*) ($r$ < 15% of its palimpsest, where $r$ means the radial great-circular distance from the center of the palimpsest), the unoriented massif facies (*umf*) (15% < $r$ <40%), the concentric massif facies (*cmf*) (40% < $r$ < 70%), and the outer deposits (*od*) (70% < $r$ <100%) (Jones et al. 2003). Beyond the radius of palimpsest, they reported that secondary craters appear. Jones et al. (2003) interpreted these zones to be (*cp*) the central plains to be formed from solidified impact melt and chunks of solid ejecta, (*umf*) a jumbled mass of ejecta located within the transient cavity, (*cmf*) a zone of ejecta-mantled preexisting crust that has been stressed and fractured by the palimpsest-producing impact, and (*od*) a continuous ejecta deposit mantling preexisting terrain. Therefore, Jones et al. (2003) concluded that the original crater rim is the boundary between the *umf* and *cmf* ($r$= 40% of the scale of palimpsest).

     Unfortunately, the Voyager or Galileo images do not show any reliable examples of secondaries or ejecta blankets that can be related to the FFI, although some of the craters in the dark terrains on Ganymede may be secondaries. Therefore, we cannot estimate the crater diameter created by

the FFI based on counting secondaries. That said, the fact that the furrows are crosscut by all recognizable impact craters may imply that the impact event would have reset Ganymede's surface completely. At least, the scale of the furrow system is roughly 4 times larger than that of the Valhalla ring system; the Valhalla ring system has the extent of 360 km < $r$ < 1900km, while the furrow system on Ganymede has the extent of 1380 km < $r$ < 7800km. If the extent of the furrow system corresponds to the *cmf* of Ganymede's palimpsests, the original crater rim would be approximately ~1380km. Therefore, the FFI must be much greater than the impactor forming Valhalla ($R_{imp} \approx 50$ km).

### 3.2. A thin plastic lithosphere model and the extent of furrows

The radial extent of Ganymede's furrows ($r$ < 7800 km) also provides an insight into the impactor size. This extent basically depends on the thickness and the strength of the satellite's lithosphere (Melosh, 1982). After (or as soon as) the formation of crater cavity, concentric faulting zone would be formed by tectonics induced by the inward flow of the asthenosphere (Melosh, 1982). We note that it is difficult to examine furrow formation itself by impact simulation due to lack of resolution. Instead, we here utilized a simple physical model known as a thin plastic lithosphere model (Melosh, 1982). Based on the model, the extent of faulting zone is a function of a single dimensionless strength parameter,

$$\gamma = \frac{\sqrt{3}Sa}{2Yt},$$

where $\gamma$ is dimensionless strength parameter, $Y$ is the strength of lithosphere, $a$ is the transient crater radius, $t$ is the thickness of lithosphere, and $S$ is the constant equal to $(0.05 - 0.1)\rho g d$, where $\rho$=900 kg/m$^3$ is the crustal density, $g$=1.428 m/s$^2$ is the gravitational acceleration at the surface of Ganymede, and $d$ is the crater depth. Larger dimensionless strength parameter (i.e. larger $\gamma$) means thinner or weaker lithosphere, which causes a greater extent of faulting zone, while smaller dimensionless strength parameter (i.e. smaller $\gamma$) means thicker or more rigid lithosphere, which forms a compact faulting zone. Following the study for Valhalla (Melosh, 1982), we assume $Y$ = 1 – 10 MPa, $S$=0.075$\rho g d$, and $d$=$a$/5. According to geomorphological analyses of furrows (Nimmo and Pappalardo, 2004), an effective elastic thickness is ~0.5 km and a brittle - ductile transition depth is ~2 to ~3 km. Thus, we assume $t$ = 0.5 – 5 km. Here we examine three cases: thin weak lithosphere model ($t$=0.5km, $Y$=1MPa), moderate thickness or weakness lithosphere ($t$=0.5km, $Y$=10MPa, or $t$=5km, $Y$=1MPa), and thick rigid lithosphere ($t$=5km, $Y$=10MPa). Following the model shown in appendix A in Melosh (1982), we calculate the extent of faulting zone as a function of transient crater radius and results are shown in Fig. 9. A transient crater with a radius of $a$ = 800km (this can be generated by $R_{imp}$ = 150 km) can result in a global-scale faulting zone even in the case of the thick and rigid lithosphere. Similarly, if we assume the thin weak lithosphere, a transient crater with $a$ > 300km is sufficient to form a global-scale faulting zone. This radius roughly corresponds to $R_{imp}$ = 50 km (equivalent to an impactor forming Valhalla).

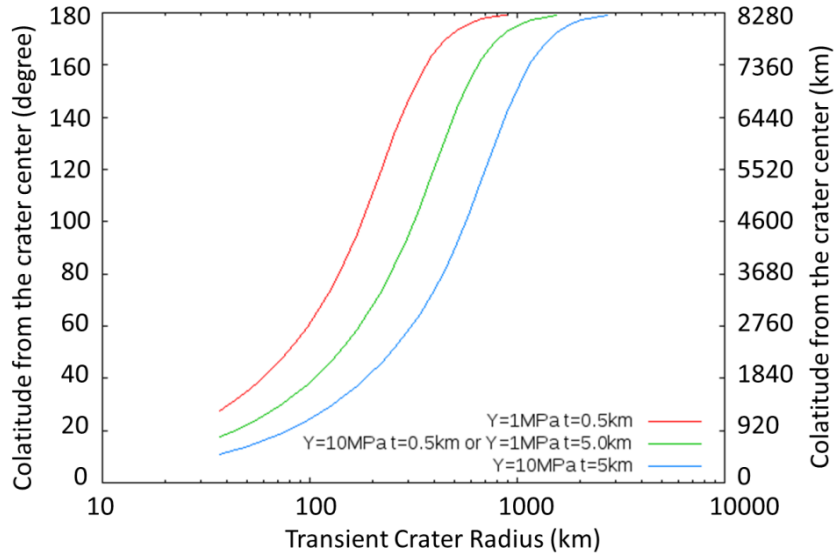

**Fig. 9.** The extent of faulting zone induced by a given transient crater radius. Vertical axis means the distance from the crater center (in degree or kilometer). Here we assume three cases: thin weak lithosphere with Y=1MPa t = 0.5km, moderate (thick weak or thin rigid) lithosphere with Y=1MPa t= 5km (or Y=10MPa t=0.5km), and thick rigid lithosphere with Y=10MPa t = 5km.

### 3.3. Numerical Simulation

Instead of aiming at reproducing the furrows themselves, we carried out numerical impact simulation by using shock-physics code iSALE (Amsden et al., 1980; Collins et al., 2004; Wünnemann et al., 2006) to estimate the size of the impactor that can reproduce the scale of the central melt pool formed by the impact. We assume an icy impactor with a radius ($R_{imp}$) of 150 km or 50km and compare the results of both cases. The detail description for the numerical simulation is given in Appendix B. We found that an impactor with $R_{imp}$ = 50 km generates a transient crater with a radius of 300km and a depth of 250 km, while an impactor with $R_{imp}$ = 150 km generates one with a radius of 800km and a depth of 1000 km (Fig. 10). In either case, materials around the impact site become fluidized and any recognizable surface structures such as rims do not form, which is consistent with the properties of the global-scale furrow system. In this simulation, an outward flow from the impact site is generated after the collapse of the central uplift and resurfaces the area around the impact site. In the case of $R_{imp}$ = 150 km, at least the hemispheric region centered at the impact site is modified. On the other hand, an impact with $R_{imp}$ = 50 km would be insufficient for global resurfacing.

    A numerical simulation to model a Valhalla-like basin using iSALE (Johnson et al., 2013) shows that an impactor with a radius of 50 km can generate a melt pool with a radius of 300 km on Callisto, and it was proposed that the melt pool would be originated from the central bright zone without ridge or trough. Similarly, our simulation shows that an impactor with $R_{imp}$ = 150 km can generate a melt pool with a radius of 1000km (here we define a melt pool by a region that experienced a temperature higher than 300K). This agrees with the observed size of the furrow-free central region of Marius Regio. On the other hand, an impactor with $R_{imp}$ = 50 km generates a small melt pool with a radius of 500 km.

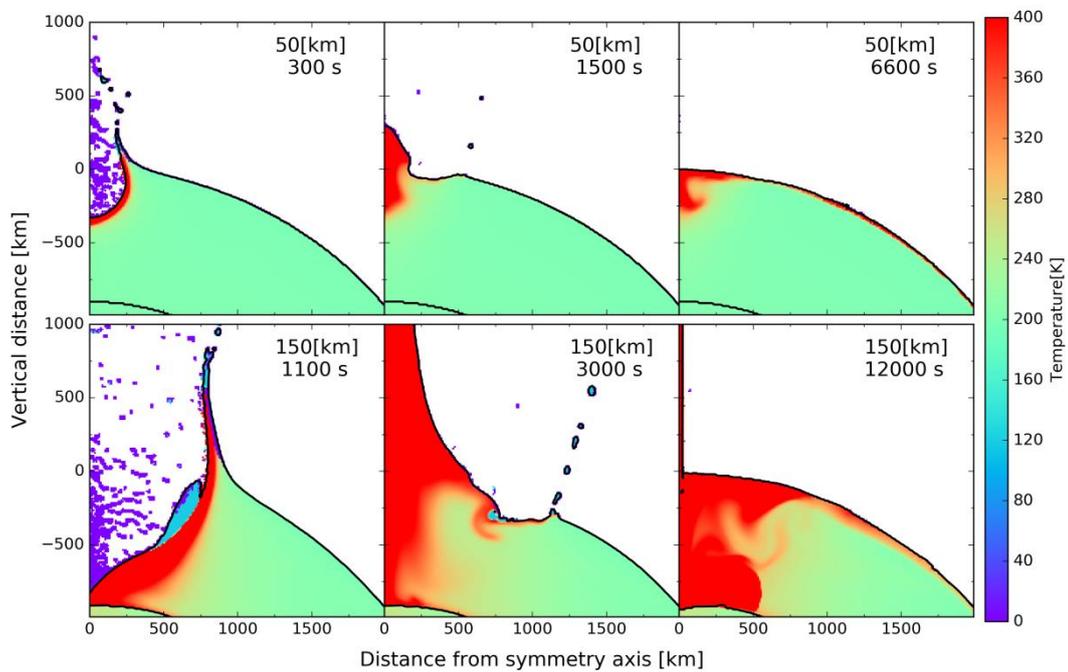

Fig. 10. Time series of simulations of head-on impacts between an impactor and Ganymede. Upper panels show the case of an impactor with a radius of 50 km (see also Movie 1 and 2), and bottom panels show the case of 150 km

(Movie 3 and 4).

### 3.4. Summary

A thin plastic lithosphere model and the extent of furrows show that an impactor with $R_{imp}$ > 50 km is able to form a global-scale faulting zone if Ganymede's lithosphere used to be thin and weak, while $R_{imp}$ = 150 km forms a global-scale faulting zone even in the case of rigid thick lithosphere. Our numerical simulation utilizing iSALE indicates that $R_{imp}$ = 150 km would resurface at least the hemispheric region around its impact site, while $R_{imp}$ = 50 km would resurface locally around the impact site. The former is consistent with the view that the furrows are crosscut by any recognizable impact craters, while the latter needs another mechanism to explain the reset of Ganymede's surface. Furthermore, an impactor with $R_{imp}$ = 150 km can generate a melt pool with a radius of 1000km, which could explain furrow-free central region of Marius Regio. Taken all together, we conclude that an impactor with a radius of 150 km is consistent with current properties of the global-scale furrow system. At least, it should be significantly greater than 50 km in radius.

### 4. Discussion and Implication

If furrows are created by an impact, this large impact should have affected the geology of Ganymede significantly. It is known that the crater density in most heavily cratered terrains of Ganymede is slightly lower than that of Callisto (Strom et al., 1981). Such old craters are considered to be formed during the Late Heavy Bombardment (LHB, 4.1~3.8 Ga), according to the so-called Nice model (Gomes et al., 2005). Perhaps, the FFI would have taken place during the LHB, would have reset the crater age of Ganymede, and would be responsible for the difference in the crater densities between Ganymede and Callisto.

Our numerical impact simulation for the case of $R_{imp}$ = 150 km ($R_{imp}$ =50 km) demonstrates the generation of a large melt region with a radius of ~1000 km (~500 km). This large melt region would allow the dense material to gravitationally segregate, which generate positive gravity anomaly such as mass concentrations seen on the lunar impact basins. Therefore, we predict that, if our view is correct, positive gravity anomaly with a radius of ~1000 km when $R_{imp}$ = 150 km (or a radius of ~500 km when $R_{imp}$ = 50 km) should exist around 20°S 180°W. This can be tested by future geodetic observations of Ganymede such as the JUICE (Jupiter Icy moon Explorer) or Europa Clipper mission. Interestingly, Ganymede is locked in synchronous rotation and the center of the furrow system coincides with the anti-Jovian longitude (180°W). This is consistent with the existence of large mass anomaly created by the FFI. Also, detailed examination of the topography may find the existence of the circular or radial structure associated with an impact basin, rim crests, or ejecta. At the same time, the question if this furrow system is really impact origin could be also tested by the future geodetic observations. For example, it is proposed that the furrow system can be formed by another mechanism, such as mantle convection (Casacchia and Strom, 1984).

The radius of the FFI is still uncertain and it may be smaller or greater than our estimate. If the FFI was twice as large as our estimate, it would have melted about half the volume of Ganymede. The accumulated dense material (iron or silicate) in the large melt region would form a large negatively buoyant mass at the bottom of the pool that stresses the underlying material and rapidly sink toward the satellite's center. This segregation can trigger the whole body differentiation (Tonks and Melosh, 1992). In contrast, Callisto does not have evidence of such large impact events. This scenario may explain the so-called Ganymede-Callisto dichotomy; Ganymede is fully differentiated while Callisto's differentiation is incomplete, despite of the similarity between Ganymede and Callisto in size and composition (Schubert et al., 2004). Note that it is proposed that Ganymede's differentiation are created by slight advantage of (i) radioactive decay and/or (ii) tidal heating, because the silicate mass fraction of Ganymede is a little larger than that of Callisto, and Ganymede is closer to the planet than Callisto (e.g. Schubert et al., 2004). Also, it is proposed that numerous impacts during the late heavy bombardment have been sufficiently energetic to lead to differentiation in Ganymede (Barr and Canup, 2010).

Although our present work demonstrated for the first time that the furrow system of Ganymede was likely created by a single large impact during LHB, it is highly desirable to derive further constraints on the origin of the furrow system from more detailed observations in addition to the geomorphological study done in the present work.

5. Conclusion

We found that almost all of the furrows over Ganymede are aligned in concentric circles centered at 20°S 180°W and the histogram of the degree of concentricity for furrows over Ganymede shows a sharp peak at 0-5° deviation from concentricity, with 85% of the ring segments aligned within 30° of concentricity. This indicates that the scale of the furrow system (so-called the Galileo Marius furrow system) on Ganymede is much greater than previously supposed, and that Ganymede used to have a global-scale multi-ring impact structure formed by a single impact event. The concentricity of the furrow system seems to suggest that a lateral movement between each dark terrain block is small or not occurred over the entire surface of Ganymede even during the formation of the bright terrains, although the surface of Ganymede has been tectonically modified significantly during formation of the bright terrains. The estimate of the impactor size is difficult, but an 150km-radius impactor is consistent with the properties of the furrow system, even though this estimate includes some uncertainty. If the above estimate is correct, we predict that Ganymede has positive gravity anomaly with a radius of ~1000 km around 20°S 180°W. These could be tested by future observations, such as JUICE or Europa Clipper mission.

Appendix A. Measurements of furrow system

We measured the geometric center of furrows and the deviation from concentricity of each furrow segment. Our measurement of Ganymede's furrow system is almost the same as the method of Schenk and McKinnon (1987). The geometric center of the furrow system was estimated by minimizing the sum of the residuals:

$\sum_i (R_a - R_b)_i^2$ , (A1)

where $R_a$ and $R_b$ were the great-circular distance between the endpoints of the $i$th furrow segment and the geometric center of the furrow system. In this paper, although the length of each furrow segment varies widely, we did not adopt weighting in each length. The deviation from concentricity of each furrow segment (angle of intersection between furrow segments and concentric circles) was defined by

$$\sin^{-1}\left(\frac{R_a - R_b}{A}\right), \quad (A2)$$

where $A$ was the length of each furrow segment (i.e. great-circular distance between $R_a$ and $R_b$).

We obtained $R_a$, $R_b$, and $A$ from furrow segments determined by Ganymede's global geologic map (Collins et al., 2013). Collins et al. (2013) identified 906 furrow segments over Ganymede: 532 in Galileo Regio, 161 in Marius, 25 in Melotte, 37 in Nicholson, 128 in Perrine, 4 in unnamed block of the dark terrain near Ta-urt crater, and 19 in the rest of Ganymede. Other than the system centered at 20°S 180°W, some distinct furrow systems exist on the dark terrains, for example, small multi-ring structures centered at 28°S 155°W (Marius Regio), 25°S 255°W (Melotte Regio), or 56°N 46°W (Perrine Regio) (e.g. Collins et al., 2013; Schenk and McKinnon, 1987). When we calculated the geometric center, we excluded furrows of these small multi-ring systems. As a result, we obtained the center of the furrow system to be 21.3°S 179.4°W based on furrow segments in Galileo and Marius Regio. Then, we excluded radial furrows (i.e. furrows whose deviation from concentricity is greater than 20°). These geometric centers obtained in the present work are consistent with the one obtained by Schenk and McKinnon (1987), 20.7°S 179.2°W. Similarly, using furrow segments over Ganymede (including Galileo, Marius, Melotte, Nicholson, Perrine Regios, and unnamed block of the dark terrain near Ta-urt crater), we obtained the center of the furrow system to be 22.9°S 178.4°W.

Furthermore, we calculated the deviation from concentricity of each furrow segment and made its histograms, assuming the center to be 21.3°S 179.4°W (Fig. 5). Note that, even if we used any of the above three centers, the histograms hardly changed. In Fig. 7, we excluded the three small multi-ring structures in Marius, Melotte, and Perrine Regios but excluded neither radial furrow systems nor the G5 multi-ring furrow system (furrows lying partly in Galileo Regio shown in G5 in Fig. 3 and Fig. 4F). Histograms for the G5 furrow system was shown in Fig. 5d. In Fig. 5a, as a comparison, we added the case of Valhalla ring system on Callisto obtained by Schenk and McKinnon (1987). Note that Schenk and McKinnon (1987) excluded radial ring segments from this histogram, and therefore it did not include ring segments with large deviation. Histogram for Perrine Regio was made from furrows in southern block of Perrine Regio shown in P1 in Fig. 3 and

Fig. 4H, because northern block of Perrine Regio includes many largely-sinuous furrows, which are probably a part of the distinct multi-ring structure.

Next, we attempt to verify quantitatively how large lateral movement between each dark terrain block can be detected by the concentricity of furrows. As an example, we shift the western portion of Nicholson Regio (approximately 40°S 40°W, N2 in Fig. 3) slightly, and see how the histogram in Fig. 5c changes. The movement of the rigid block on a spherical surface can be described as a rotation about a fixed axis known as an Euler pole. The change of the histograms after the movement largely depends on the location of the pole. For example, if the Euler pole of the movement of Nicholson Regio is 20°S 180°W, the histogram of the deviation does not change at all. In this section, we tested 15 simple cases: 10°, 20°, and 30° movement toward north, south, west, and east and 10°, 20°, and 30° right-hand thread rotation at current location without shift. When moving Nicholson toward north, we assumed the Euler pole of the rotation to be 0°N 130°W. When rotating Nicholson without shift, we assumed the Euler pole of the rotation to be 40°S 40°W. Then, Rodriguez's rotation formula was used to calculate these movements. The operation that all of furrows are moved by the formula is actually equal to one that the geometric center of 21.3°S 179.4°W is moved by the inverse formula. Therefore, we calculated angle of intersection between furrow segments at current location and the concentric circles centered at the center that was moved inversely. The results were shown in Fig. 7.

## Appendix B. Numerical Method

We use the iSALE-2D shock physics code (Wünnemann et al., 2006), which is an extension of the SALE hydrocode (Amsden et al., 1980). To simulate hypervelocity impact processes in solid materials, SALE was modified to include an elasto-plastic constitutive model, fragmentation models, various equations of state, and multiple materials (Ivanov et al., 1997; Melosh et al., 1992). More recent improvements include a modified strength model (Collins et al., 2004), a porosity compaction model (Collins et al., 2011; Wünnemann et al., 2006) and a dilatancy model (Collins, 2014).

We employ the two-dimensional cylindrical coordinate system and perform head-on impact simulations between Ganymede and an impactor.

We assume that Ganymede is differentiated into an icy mantle and a rocky core. Its radius and the thickness of the icy mantle are fixed at 2600 km and 900 km, respectively (Schubert et al., 2004). We performed simulations with several different sizes of the core, and confirmed that the size of the melt pool does not change significantly. For example, Figure A1 shows the temperature distribution in a coreless Ganymede after the impact. The size of the melt pool was roughly 1000 km, which was consistent with the differentiated Ganymede case (Fig.10). The impactor is assumed to be composed of ice, and its impact velocity is fixed at 20 km/s (Zahnle et al., 2003). We assume an impactor radius ($R_{imp}$) to be 50 km or 150 km and compare the results of both cases. Following previous works, we use the Tillotson equation of state for ice (Tillotson, 1962) and ANEOS equation of state for dunite (Thompson and Lauson, 1974). The strength and damage model parameters for ice and dunite are listed in Table A1, whose values are used by previous works (Johnson et al., 2016a; Johnson et al., 2016b; Senft and Stewart, 2008). We assume the temperature of the surface and icy mantle of the target to be 120 K and 200-210 K, respectively (Nimmo and Pappalardo, 2004; Schubert et al., 2004). We assume the temperature of the impactor to be equal to the surface of Ganymede, 120 K. We also consider the effect of acoustics fluidization using results obtained by impact simulations (Bray et al., 2014). We assume that the angle of impact is perpendicular to the horizon. Note that an oblique impact cannot be simulated in iSALE-2D. Most of grid-based hydrodynamic codes cannot treat the undifferentiated body with the mixture of rock and ice, and therefore, ice/ice impacts have been used to simulate impacts onto undifferentiated bodies. This assumption has been considered to be reasonable because rock particles hardly influence the ice EOS when the size of the rock particle embedded in the ice is small (e.g. Bar and Canup, 2011).

Since we aim to evaluate the size of a melt pool after the impact in this impact simulation, we perform numerical integration for a period of time that is sufficiently long for the impact site to become stabilized. When $R_{imp}$ = 50 km, the required simulation period is $T$=5000-10000 s, while it is $T$=10000-15000 s when $R_{imp}$=150 km. In order to examine the extent of ejecta blankets, we carry out impact simulation using large computational domain of 7800 km × 13000 km, whose spatial cell size is 10 km. The iSALE-2D supports two types of gravity field model: central gravity and self-gravity. The central gravity model can calculate spatially varying gravity field, which is independent of time. On the other hand, the gravity field by the self-gravity model is calculated from mutual gravity between the computational cells, and is updated periodically (Davison et al., 2012). In this study, the temporal variation of the gravity field can be assumed to be small because the target deformation due to the impact is small. Thus, we apply the central gravity model. Indeed, although we also carried out impact simulations using the self-gravity model, outcomes were almost the same as the case of the central gravity model (Fig. A2).

Table A1. iSALE material parameters

| Description | Ice | Dunite |
| --- | --- | --- |
| Poisson's ratio | 0.33 [*1] | 0.25 [*5] |
| Thermal softening | 1.84 [*1] | 1.1 [*5] |
| Melting temperature (K) | 273.15 [*1] | 1373 [*5] |
| Simon A parameter (Pa) | ----- | $1.52 \times 10^9$ [*5] |
| Simon C parameter | ----- | 4.05 [*5] |
| Strength model | ICE [*2] | ROCK [*5] |
| Cohesion (undamaged) (Pa) | $1.64 \times 10^7$ [*1] | $5.07 \times 10^6$ [*5] |
| Coefficient of internal friction (undamaged) | 6.54 [*1] | 1.58 [*5] |
| Limiting strength at high pressure (Pa) | $1.47 \times 10^8$ [*1] | $3.26 \times 10^9$ [*5] |
| Cohesion (damaged) (Pa) | $10^4$ [*1] | $10^4$ [*5] |
| Coefficient of internal friction (damaged) | 0.55 [*1] | 0.63 [*5] |
| Limiting strength at high pressure (Pa) | $1.47 \times 10^8$ [*1] | $3.26 \times 10^9$ [*5] |
| Damage model | COLLINS [*6] | COLLINS [*6] |
| Brittle ductile transition (Pa) | $6.89 \times 10^8$ [*1] | $1.23 \times 10^9$ [*5] |
| Brittle plastic transition (Pa) | $6.99 \times 10^8$ [*1] | $2.35 \times 10^9$ [*5] |
| Tensile strength (Pa) | $1.7 \times 10^5$ [*1] | $10^7$ [*5] |
| Acoustic fluidization model | Block [*3] | ----- |
| Acoustic fluidization viscosity constant | 0.553 ($R_{imp}$=50km) 0.785 ($R_{imp}$=150km) [*4] | ----- |
| Acoustic fluidization decay time constant | 45.1 ($R_{imp}$=50km) 24.9 ($R_{imp}$=150km) [*4] | ----- |

*1 (Johnson et al., 2016b; Senft and Stewart, 2008)
*2 (Collins et al., 2016)
*3 (Ivanov and Kostuchenko, 1997)
*4 (Bray et al., 2014)
*5 (Johnson et al., 2016a)
*6 (Collins et al., 2004)

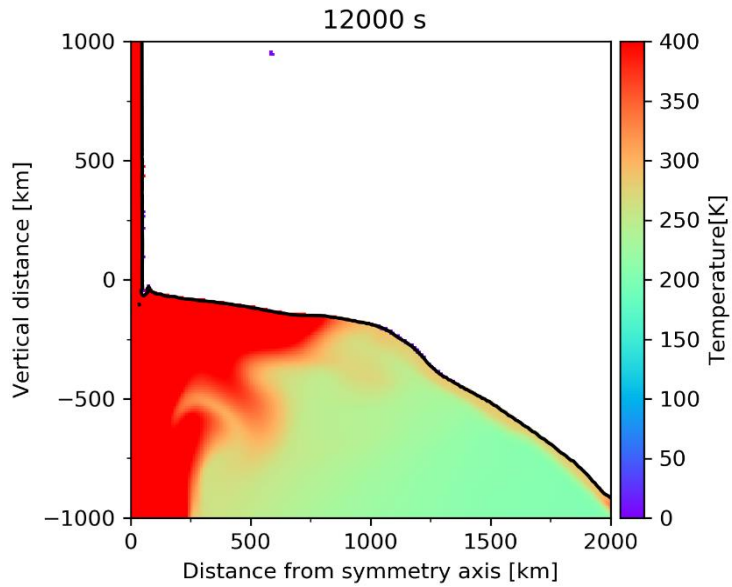

**Figure A1.** The size of melt pool for the case of coreless Ganymede.

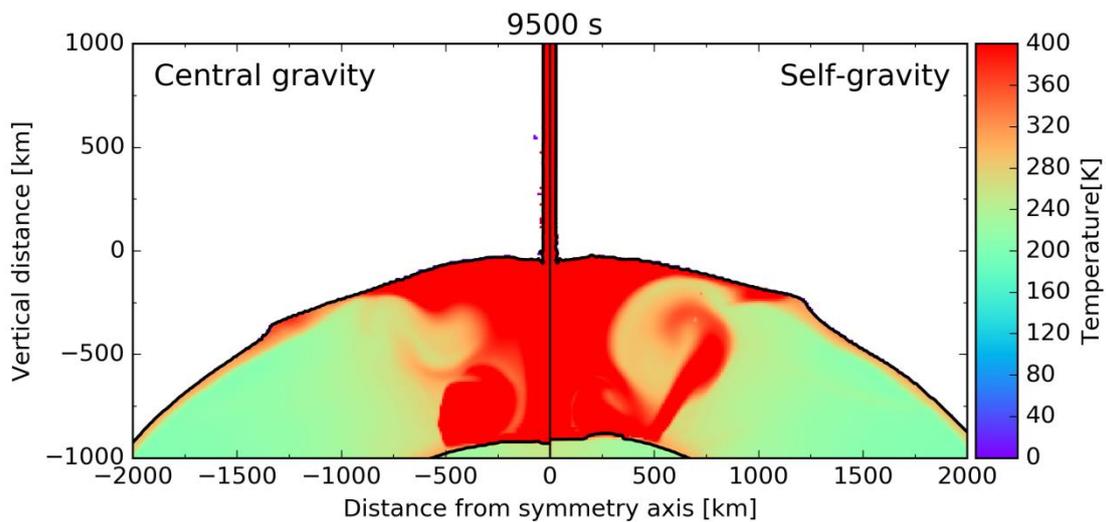

**Figure A2.** The size of melt pool on central gravity model (left) and self-gravity model (right) in the case of $R_{\mathrm{imp}}$=150 km (see also Movie 5).

Acknowledgements

The authors wish to thank Geoffrey Collins for his comments, which significantly tightened the manuscript. This work was partly supported by JSPS KAKENHI Grant Number JP15H03716 (all), JP19K14787 (R.S.), 20K14538, and 20H04614 (N.H.) and by the Hyogo Science and Technology Association (N.H.). Numerical calculations were in part performed using the PC cluster at the National Astronomical Observatory of Japan. We gratefully acknowledge the developers of iSALE, including Gareth Collins, Kai Wünnemann, Boris Ivanov, H. Jay Melosh, Dirk Elbeshausen, and Thomas Davison.

## Supporting information

There are one supplementary figure and five movies. Captions for Figure S1 and Movies are listed below.

Figure S1. The original-sized image of Fig. 3.

Movie 1. The animation of upper panels of Fig. 10.

Movie 2. The entire of Ganymede of Movie 1.

Movie 3. The animation of bottom panels of Fig. 10.

Movie 4. The entire of Ganymede of Movie 3.

Movie 5. The animation of Figure A2.

## References

Amsden, A. A., H. M. Ruppel, and C. W. Hirt (1980), SALE: a simplified ALE computer program for fluid flow at all speeds, Los Alamos National Laboratory Report LA-8095 – 1980, Los Alamos, NM.

Barr, A. C., R. M. Canup, (2010), Origin of the Ganymede-Callisto dichotomy by impacts during the late heavy bombardment, *Nature Geoscience 3*, 164.

Bray, V. J., G. S. Collins, J. V. Morgan, H. J. Melosh, and P. M. Schenk (2014), Hydrocode simulation of Ganymede and Europa cratering trends – How thick is Europa's crust?, *Icarus, 231*, 394-406, doi:https://doi.org/10.1016/j.icarus.2013.12.009.

Casacchia, R., and R. G. Strom (1984), Geologic evolution of Galileo Regio, Ganymede, *Journal of Geophysical Research: Solid Earth*, *89*(S02), B419-B428, doi:10.1029/JB089iS02p0B419.

Collins, G. C., G. W. Patterson, J. W. Head, R. T. Pappalardo, L. M. Prockter, B. K. Lucchitta, and J. P. Kay (2013), Global geologic map of Ganymede: U.S. Geological Survey Scientific Investigations Map 3237, pamphlet 4 p., 1 sheet, scale 1:15,000,000,

http://dx.doi.org/10.3133/sim3237.

Collins, G. S. (2014), Numerical simulations of impact crater formation with dilatancy, *Journal of Geophysical Research: Planets*, *119*(12), 2600-2619, doi:doi:10.1002/2014JE004708.

Collins, G. S., D. Elbeshausen, K. Wünnemann, T. M. Davison, B. A. Ivanov, and H. J. Melosh (2016), iSALE: A multi-material, multi-rheology shock physics code for simulating impact phenomena in two and three dimensions. iSALE-Dellen manual.

Collins, G. S., H. J. Melosh, and K. Wünnemann (2011), Improvements to the ε-α porous compaction model for simulating impacts into high-porosity solar system objects, *International Journal of Impact Engineering*, *38*(6), 434-439, doi:https://doi.org/10.1016/j.ijimpeng.2010.10.013.

Collins, G. S., M. H. Jay, and I. B. A. (2004), Modeling damage and deformation in impact simulations, *Meteoritics & Planetary Science*, *39*(2), 217-231, doi:doi:10.1111/j.1945-5100.2004.tb00337.x.

Davison, T. M., F. J. Ciesla, and G. S. Collins (2012), Post-impact thermal evolution of porous planetesimals, *Geochimica et Cosmochimica Acta*, *95*, 252-269, doi:https://doi.org/10.1016/j.gca.2012.08.001.

Frieden, B. R., and W. Swindell (1976), Restored Pictures of Ganymede, Moon of Jupiter, *Science*, *191*(4233), 1237-1241, doi:10.1126/science.191.4233.1237.

Gomes, R., H. F. Levison, K. Tsiganis, and A. Morbidelli (2005), Origin of the cataclysmic Late Heavy Bombardment period of the terrestrial planets, *Nature*, *435*, 466, doi:10.1038/nature03676.

Greeley, R., J. E. Klemaszewski, and R. Wagner (2000), Galileo views of the geology of Callisto, *Planetary and Space Science*, *48*(9), 829-853, doi:https://doi.org/10.1016/S0032-0633(00)00050-7.

Ivanov, B. A., D. Deniem, and G. Neukum (1997), Implementation of dynamic strength models into 2D hydrocodes: Applications for atmospheric breakup and impact cratering, *International Journal of Impact Engineering*, *20*(1), 411-430, doi:https://doi.org/10.1016/S0734-743X(97)87511-2.

Ivanov, B., and V. Kostuchenko (1997), Block oscillation model for impact crater collapse, in *28th Lunar and Planetary Science Conference*, Abstract #631, Lunar and Planetary Institute, Houston.

Johnson, B. C., T. J. Bowling, A. J. Trowbridge, and A. M. Freed (2016b), Formation of the Sputnik Planum basin and the thickness of Pluto's subsurface ocean, *Geophysical Research Letters*, *43*(19), 10,068-10,077, doi:doi:10.1002/2016GL070694.

Johnson, B. C., T. J. Bowling, and H. J. Melosh (2013), Formation of Valhalla-Like

Multi-Ring Basins, in *44th Lunar and Planetary Science Conference*, Abstract #1302, Lunar and Planetary Institute, Houston.

Johnson, B. C., et al. (2016a), Formation of the Orientale lunar multiring basin, *Science*, *354*(6311), 441-444, doi:10.1126/science.aag0518.

Jones, K. B., J. W. Head III, R. T. Pappalardo, J. M. Moore, (2003), Morphology and origin of palimpsests on Ganymede based on Galileo observations, *Icarus 164(1)*, 197-212.

McKinnon, W. B., and E. Parmentier (1986), Ganymede and Callisto, in *Satellites* (Eds. J. A. Burns and M. S. Matthews), University of Arizona Press, Tucson.

McKinnon, W. B., and H. J. Melosh (1980), Evolution of planetary lithospheres: Evidence from multiringed structures on Ganymede and Callisto, *Icarus*, *44*(2), 454-471, doi:http://dx.doi.org/10.1016/0019-1035(80)90037-8.

Melosh, H. J. (1982), A simple mechanical model of Valhalla Basin, Callisto, *Journal of Geophysical Research: Solid Earth*, *87*(B3), 1880-1890, doi:10.1029/JB087iB03p01880.

Melosh, H. J., E. V. Ryan, and E. Asphaug (1992), Dynamic fragmentation in impacts: Hydrocode simulation of laboratory impacts, *Journal of Geophysical Research: Planets*, *97*(E9), 14735-14759, doi:10.1029/92JE01632.

Nimmo, F., and R. T. Pappalardo (2004), Furrow flexure and ancient heat flux on Ganymede, *Geophysical Research Letters*, *31*(19), L19701, doi:10.1029/2004GL020763.

Pappalardo, R. T., G. C. Collins, J. Head, P. Helfenstein, T. B. McCord, J. M. Moore, L. M. Prockter, P. M. Schenk, and J. R. Spencer (2004), Geology of Ganymede, in *Jupiter: The Planet, Satellites and Magnetosphere* (Eds. F. Bagenal et al.), Cambridge University Press, Cambridge, United Kingdom.

Passey, Q. R., and E. M. Shoemaker (1982), Craters and basins on Ganymede and Callisto-Morphological indicators of crustal evolution, in *Satellites of Jupiter* (Eds. D. Morrison and M. S. Matthews), University of Arizona Press, Tucson.

Prockter, L. M., G. C. Collins, S. L. Murchie, P. M. Schenk, and R. T. Pappalardo (2002), Ganymede Furrow Systems as Strain Markers: Implications for Evolution and Resurfacing Processes, in *33rd Lunar and Planetary Science Conference*, Abstract #1272, Lunar and Planetary Institute, Houston.

Schenk, P. M. (1995), The geology of Callisto, *Journal of Geophysical Research: Planets*, *100*(E9), 19023-19040, doi:10.1029/95JE01855.

Schenk, P. M., and F. J. Ridolfi (2002), Morphology and scaling of ejecta deposits on icy satellites, *Geophysical Research Letters*, *29*(12), 1590, doi:10.1029/2001GL013512.

Schenk, P. M., and W. B. McKinnon (1987), Ring geometry on Ganymede and Callisto, *Icarus*, *72*(1), 209-234, doi:http://dx.doi.org/10.1016/0019-1035(87)90126-6.

Schubert, G., J. Anderson, T. Spohn, and W. McKinnon (2004), Interior composition,

structure and dynamics of the Galilean satellites, in *Jupiter: The Planet, Satellites and Magnetosphere* (Eds. F. Bagenal et al.), Cambridge University Press, Cambridge, United Kingdom.

Senft, L. E., and S. T. Stewart (2008), Impact crater formation in icy layered terrains on Mars, *Meteoritics & Planetary Science*, *43*(12), 1993, doi:doi:10.1111/j.1945-5100.2008.tb00657.x.

Shoemaker, E. M., and R. Wolfe (1982), Cratering time scales for the Galilean satellites, in *Satellites of Jupiter* (Eds. D. Morrison and M. S. Matthews), University of Arizona Press, Tucson.

Smith, B. A., et al. (1979a), The Galilean Satellites and Jupiter: Voyager 2 Imaging Science Results, *Science*, *206*(4421), 927-950, doi:10.1126/science.206.4421.927.

Smith, B. A., et al. (1979b), The Jupiter System Through the Eyes of Voyager 1, *Science*, *204*(4396), 951-972, doi:10.1126/science.204.4396.951.

Strom, R. G., A. Woronow, and M. Gurnis (1981), Crater populations on Ganymede and Callisto, *Journal of Geophysical Research: Space Physics*, *86*(A10), 8659-8674, doi:doi:10.1029/JA086iA10p08659.

Thompson, S., and H. Lauson (1974), Improvements in the Chart D radiation-hydrodynamic CODE III: Revised analytic equations of state, Sandia National Laboratories Report SC-RR--71-0714, Albuquerque, NM.

Tillotson, J. H. (1962), Metallic equations of state for hypervelocity impact, General Atomic Report GA-3216, Division of General Dynamics, John Jay Hopkins Laboratory for Pure and Applied Science, San Diego, CA.

Tonks, W. B., and H. Jay Melosh (1992), Core formation by giant impacts, *Icarus*, *100*(2), 326-346, doi:https://doi.org/10.1016/0019-1035(92)90104-F.

U.S. Geological Survey (2003), Controlled color photomosaic map of Ganymede; Jg 15M CMNK: U.S. Geological Survey Geologic Investigations Series I–2762, available at https://pubs.usgs.gov/imap/i2762/.

Wünnemann, K., G. S. Collins, and H. J. Melosh (2006), A strain-based porosity model for use in hydrocode simulations of impacts and implications for transient crater growth in porous targets, *Icarus*, *180*(2), 514-527, doi:https://doi.org/10.1016/j.icarus.2005.10.013.

Zahnle, K., P. Schenk, H. Levison, and L. Dones (2003), Cratering rates in the outer Solar System, *Icarus*, *163*(2), 263-289.

Zuber, M. T., and E. M. Parmentier (1984), A geometric analysis of surface deformation: Implications for the tectonic evolution of Ganymede, *Icarus*, *60*(1), 200-210, doi:http://dx.doi.org/10.1016/0019-1035(84)90148-9.